\documentclass[journal]{IEEEtran}
\usepackage{cite}
\usepackage{graphicx}
\usepackage{amssymb,amsmath,bm}
\usepackage[colorlinks,linkcolor=black,anchorcolor=black,citecolor=black,urlcolor=black]{hyperref}
\usepackage{hyperref}
\usepackage{multirow}
\usepackage{amsfonts}
\usepackage{url}
\usepackage{multirow}
\usepackage{color}
\usepackage{tabularx}
\usepackage{color}
\usepackage{diagbox}
\ifCLASSINFOpdf
\else
\fi
\begin{document}
%
\title{APCodec: A Neural Audio Codec with Parallel Amplitude and Phase Spectrum Encoding and Decoding}
%
%
%

\author{Yang~Ai,~\IEEEmembership{Member,~IEEE}, Xiao-Hang Jiang, Ye-Xin Lu, Hui-Peng Du,~Zhen-Hua~Ling,~\IEEEmembership{Senior Member,~IEEE}
\thanks{This work was funded by the National Nature Science Foundation of China under Grant 62301521 and U23B2053, the Anhui Provincial Natural Science Foundation under Grant 2308085QF200, and the Fundamental Research Funds for the Central Universities under Grant WK2100000033.}
\thanks{Y. Ai, X.-H. Jiang, Y.-X. Lu, H.-P. Du and Z.-H. Ling are with the National Engineering Research Center of Speech and Language Information Processing, University of Science and Technology of China, Hefei, 230027, China (e-mail: yangai@ustc.edu.cn, jiang\_xiaohang@mail.ustc.edu.cn, yxlu0102@mail.ustc.edu.cn, redmist@mail.ustc.edu.cn, zhling@ustc.edu.cn).}
\thanks{Corresponding author: Zhen-Hua Ling.}
}
%
%

\markboth{}%
{Shell \MakeLowercase{\textit{et al.}}: Bare Demo of IEEEtran.cls for Journals}
%



\maketitle

\begin{abstract}
This paper introduces a novel neural audio codec targeting high waveform sampling rates and low bitrates named APCodec, which seamlessly integrates the strengths of parametric codecs and waveform codecs.
The APCodec revolutionizes the process of audio encoding and decoding by concurrently handling the amplitude and phase spectra as audio parametric characteristics like parametric codecs.
It is composed of an encoder and a decoder with the modified ConvNeXt v2 network as the backbone, connected by a quantizer based on the residual vector quantization (RVQ) mechanism.
The encoder compresses the audio amplitude and phase spectra in parallel, amalgamating them into a continuous latent code at a reduced temporal resolution.
This code is subsequently quantized by the quantizer.
Ultimately, the decoder reconstructs the audio amplitude and phase spectra in parallel, and the decoded waveform is obtained by inverse short-time Fourier transform.
To ensure the fidelity of decoded audio like waveform codecs, spectral-level loss, quantization loss, and generative adversarial network (GAN) based loss are collectively employed for training the APCodec.
To support low-latency streamable inference, we employ feed-forward layers and causal deconvolutional layers in APCodec, incorporating a knowledge distillation training strategy to enhance the quality of decoded audio.
Experimental results confirm that our proposed APCodec can encode 48 kHz audio at bitrate of just 6 kbps, with no significant degradation in the quality of the decoded audio.
At the same bitrate, our proposed APCodec also demonstrates superior decoded audio quality and faster generation speed compared to well-known codecs, such as Encodec, AudioDec and DAC.

\end{abstract}


%
\IEEEpeerreviewmaketitle

\section{Introduction}

\IEEEPARstart{A}
{udio} codec, an important signal processing technique, compresses audio signals into discrete codes and then uses these codes to reconstruct the original audio.
In general, an encoder, a quantizer, and a decoder are the three main components of an audio codec.
The purpose of an audio codec is to utilize as few bits as possible (i.e., low bitrate) to store or transmit an audio signal while ensuring that the decoded audio quality doesn't undergo significant degradation.
Audio codec technology holds a central position in fields such as audio communication and transmission \cite{brandenburg1994iso,tremain1976linear,kroon1986regular,salami1994toll}.
Recently, audio codec technology has also been gradually applied to some downstream tasks.
For example, some researchers use the discrete codes generated by audio codecs as intermediate representations, combined with large language model technology, to achieve impressive zero-shot text-to-speech (TTS) results \cite{borsos2023audiolm,wang2023neural,zhang2023speechtokenizer,huang2023repcodec,ren2023fewer}.

Audio codec has several key properties, which are also important metrics for evaluating the audio codec:
1) Decoded audio quality, reflecting the ability of an audio codec to restore compressed audio with as minimal loss as possible.
2) Bitrate, representing compression efficiency, indicating how many bits are used to represent the discrete codes generated by the audio codec.
3) Generation speed, denoting the overall running efficiency of audio encoding, quantization, and decoding.
4) Latency, a strict requirement for real-time audio communication, indicating the minimum amount of time needed for the codec to initiate its operations.
In general, an audio codec that offers high decoded audio quality, low bitrate, fast generation speed and low latency is essential for applications such as audio communication.
However, certain applications in downstream tasks typically prioritize the decoded audio quality over latency considerations. Factoring in latency can often have an impact on the overall results.

Audio codecs are generally categorized into two main types: parametric codecs and waveform codecs.
The parametric codecs treat the characteristic parameters of audio signals as the objects for encoding and decoding, such as linear predictive coding (LPC) \cite{o1988linear}, Opus \cite{valin2013high} and EVS \cite{dietz2015overview}.
Due to the short-term stationary nature of audio signals, the characteristic parameters are frame-level and have a low update frequency.
Hence, the advantage of parametric codecs lies in their low bitrate.
However, the drawback of such codecs is their poor decoded audio quality and susceptibility to noise.
With the advancement of deep learning, researchers have employed neural vocoders to transform encoded characteristic parameters into audio waveforms, aiming to enhance the quality of decoded audio \cite{kleijn2018wavenet,klejsa2019high,valin2019real,mustafa2021streamwise,zheng2023cqnv}.
Recently, some approaches encode and decode the modified discrete cosine transform (MDCT) spectrum using neural networks, ultimately restoring the audio waveform through the inverse MDCT process \cite{davidson2023high,lim2023end}.
Unfortunately, as reported in \cite{davidson2023high,lim2023end}, these MDCT-based approaches typically necessitate high bitrate ($>$20 kbps at sampling rate of 48 kHz), thereby conflicting with the benefits of parametric codecs.

The waveform codecs aim at encoding the input audio waveform directly and reproducing a faithful reconstruction of the input audio waveform, such as pulse code modulation (PCM) \cite{black1947pulse}.
Although waveform codecs can decode high-quality audio, they also require a higher bitrate, which increases storage and transmission costs.
In recent times, end-to-end neural waveform codecs with raw waveform I/O have surfaced, offering a partial equilibrium between decoded audio quality and bitrate \cite{kankanahalli2018end,van2017neural,garbacea2019low,zhen2019cascaded,zeghidour2021soundstream,defossez2022high}.
For example, SoundStream \cite{zeghidour2021soundstream} and Encodec \cite{defossez2022high} employed the residual vector quantization (RVQ) mechanism \cite{vasuki2006review} to reduce the bitrate, while utilizing the losses from the HiFi-GAN vocoder \cite{kong2020hifi} to ensure the fidelity of the decoded audio.
Recently, researchers have made improvements addressing the issues present in current end-to-end audio codecs, primarily focusing on quantization strategies.
On the one hand, in applications such as audio communication, audio codecs have incorporated variations of RVQ to decrease bitrates and improve communication efficiency \cite{yang2023hifi,xu2023intra,jenrungrot2023lmcodec}.
For example, HiFi-Codec \cite{yang2023hifi} has introduced group RVQ (GRVQ) to reduce information redundancy in RVQ.
This allows for high-quality audio coding with less codebooks, resulting in a reduced bitrate.
On the other hand, in downstream tasks such as TTS, efforts have been undertaken to introduce or disentangle semantic information during the quantization stage, aiming to tailor the approach to the specific tasks \cite{zhang2023speechtokenizer,huang2023repcodec,ren2023fewer}.
Moreover, there have been endeavors to improve the model structure \cite{xu2023intra} or incorporate additional signal processing techniques (e.g., bandwidth extension) in codecs \cite{xiao2023multi}.
Although these codecs have indeed enhanced the decoded audio quality and decreased bitrates, they still required more than a hundred times the downsampling and upsampling operations due to the direct waveform encoding and decoding, leading to high model complexity.
Besides, direct waveform encoding and decoding could also potentially result in low generation efficiency.
Some recent works have also overlooked considerations for low latency, making it challenging to achieve streamable inference \cite{yang2023hifi,xu2023intra}.


Beyond the aforementioned challenges, encompassing bitrate, generation speed, and latency in existing audio codecs, there is presently scant research devoted to audio codecs tailored for higher waveform sampling rates (e.g., 48 kHz).
Currently, neural audio codecs (e.g., SoundStream \cite{zeghidour2021soundstream} and HiFi-Codec \cite{yang2023hifi}) are mostly designed for processing audio at sampling rates of 16 kHz or 24 kHz.
This limitation hinders the utilization of audio codecs for compressing high-sampling-rate audio data and poses challenges for downstream tasks like TTS, which aim to meet the demand for higher quality speech generation.
The aforementioned MDCT-based parametric codecs \cite{davidson2023high,lim2023end}, while targeted at 48 kHz audio, demands an excessively high bitrate.
Although AudioDec \cite{wu2023audiodec} can achieve 48 kHz audio codec at bitrate of 12.8 kbps, it still requires the integration of a neural vocoder and the adoption of a multi-stage training strategy, as reported in \cite{wu2023audiodec}.
Descript audio codec (DAC) \cite{kumar2024high} can achieve 44.1 kHz audio coding at a bitrate of only 8 kbps, thanks to improved RVQ for improving codebook utilization and improved losses to enhance the decoded audio quality.
However, the DAC's bitrate remains relatively high and lacks consideration for low latency.

To address the aforementioned challenges, this paper proposes a novel neural audio codec named APCodec.
It endeavors to provide high-quality decoded audio, while maintaining a low bitrate, fast generation speed, and low latency, specifically tailored for 48 kHz audio.
Like parametric codecs, the proposed APCodec regards amplitude and phase spectra as audio parametric characteristics during the encoding and decoding processes, rather than directly processing the raw waveform.
A notable advantage of this approach lies in its simplicity, as it only requires uncomplicated downsampling to obtain latent codes at an appropriately low sampling rate, thereby effectively reducing the bitrate.
RVQ \cite{vasuki2006review} is also utilized for code quantization to further reduce the bitrate.
With the objective of achieving faithful waveform reconstruction akin to waveform codecs, a comprehensive combination of spectral-level loss, quantization loss and generative adversarial network (GAN) based loss are employed to train the APCodec.
To attain streamable inference, a low-latency implementation is achieved by integrating feed-forward layers and causal deconvolutional layers, complemented by the application of a knowledge distillation training strategy.
The resulting fixed latency is only 6.67 ms for the 48 kHz audio codec.
Experimental results have confirmed that the proposed APCodec can achieve high-quality 48 kHz audio coding at a bitrate of only 6 kbps with only 8$\times$ downsampling/upsampling.
At the same bitrate, our proposed APCodec significantly outperforms several well-known neural codecs which support high-sampling-rate audio coding, e.g., Encodec \cite{defossez2022high}, AudioDec \cite{wu2023audiodec} and DAC \cite{kumar2024high} in terms of decoded audio quality.
The APCodec also demonstrates the fastest generation speed, attaining an impressive 89$\times$ real-time performance on GPU and 5.8$\times$ real-time performance on CPU.
This remarkable acceleration is attributed to its comprehensive all-frame-level processing.

There are three main contributions of the proposed APCodec.
Firstly, the APCodec targets audio encoding and decoding at high sampling rates and low bitrates, meeting the demands for high-sampling-rate audio compression and generation.
Secondly, the APCodec utilizes amplitude and phase spectra as the encoding and decoding entities, rather than waveforms, thereby further enhancing generation efficiency.
Thirdly, the APCodec introduces knowledge distillation to enhance the effectiveness of causal audio codec models.
This approach provides valuable insights into realizing low-latency implementations in contemporary audio codec technology.

This paper is organized as follows:
In Section \ref{sec: Proposed Methods}, we provide details on our proposed APCodec.
In Section \ref{sec: Experiments}, we present our experimental results.
Finally, we give conclusions in Section \ref{sec: Conclusion}.

\begin{figure*}
    \centering
    \includegraphics[width=\textwidth]{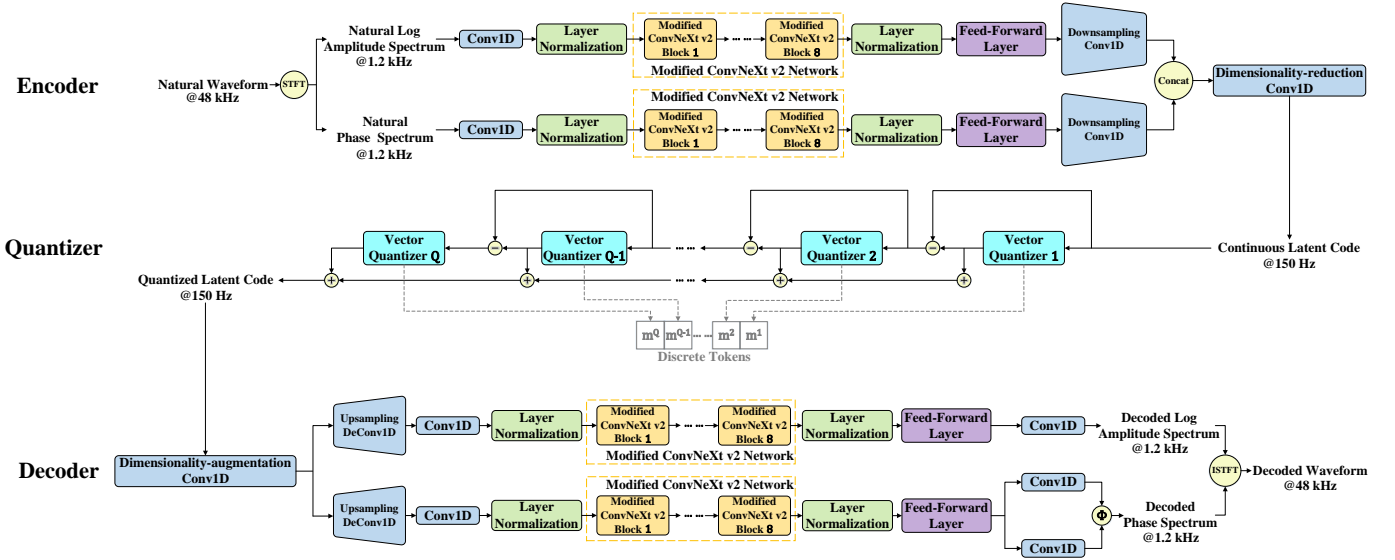}
    \caption{Details of the model structure of the proposed APCodec. Here, \emph{Conv1D}, \emph{DeConv1D}, \emph{Concat}, \emph{$\Phi$}, \emph{STFT} and \emph{ISTFT} represent the 1D convolutional layer, 1D deconvolutional layer, concatenation, phase calculation formula, short-time Fourier transform and inverse short-time Fourier transform, respectively. For waveforms, the content after @ represents the sampling rate, while for spectra and codes, the content after @ represents the frame rate (taking a sampling rate of 48 kHz and a bitrate of 6 kbps as an example).
    }
    \label{fig: APCodec}
\end{figure*}

\section{Proposed Methods}
\label{sec: Proposed Methods}

Unlike some well-known neural waveform codecs, e.g., SoundStream \cite{zeghidour2021soundstream}, Encodec \cite{defossez2022high}, HiFi-Codec \cite{yang2023hifi}, AudioDec \cite{wu2023audiodec} and DAC \cite{kumar2024high}, our proposed APCodec encodes and quantizes amplitude and phase spectra extracted from the audio waveform through short-time Fourier transform (STFT).
Finally, it decodes the quantized codes into amplitude and phase spectra and restores the audio waveform through inverse STFT (ISTFT).
Subsequently, we will present a detailed overview of the model structure and training criteria of the proposed APCodec.
Additionally, we will discuss the low-latency implementation for APCodec.

\subsection{Model Structure}
\label{subsec: Model Structure}

As illustrated in Fig. \ref{fig: APCodec}, the proposed APCodec is consist of an encoder, a quantizer and a decoder.
The APCodec utilizes amplitude and phase spectra as audio parametric characteristics for encoding and decoding, incorporating the advantages of parametric codecs to reduce bitrates.
The specific structures of these three components are outlined as follows.

\subsubsection{Encoder}
\label{subsubsec: Encoder}

As illustrated in Fig. \ref{fig: APCodec}, the encoder takes the log amplitude spectrum $\bm{A}\in \mathbb{R}^{F\times N}$ and phase spectrum $\bm{P}\in \mathbb{R}^{F\times N}$ extracted from the audio waveform $\bm{x}\in \mathbb{R}^T$ using STFT as inputs and encodes them into a continuous latent code $\bm{C}\in \mathbb{R}^{F_c\times N_c}$ that contains fused amplitude and phase information in parallel.
Here, $T$ represents the number of time-domain waveform samples, and $F$ and $N$ respectively represent the number of spectral frames and frequency bins.
Assuming the sampling rate of $\bm{x}$ is $f_s$ and the frame shift of the STFT is $w_s$, the resulting frame rate for the extracted amplitude and phase spectra is $f_s/w_s$, and it holds that $T=F\cdot w_s$.
$F_c$ and $N_c$ denote the number of frames and dimensionality of the code, respectively.

The encoder comprises parallel amplitude and phase sub-encoders that share an identical network architecture, as shown in Fig. \ref{fig: APCodec}.
In the amplitude/phase sub-encoder, the input log amplitude/phase spectrum is initially processed through an input 1D convolutional layer (channel size $=K$), a layer normalization \cite{ba2016layer} and then undergoes deep processing through a modified ConvNeXt v2 network.
The output of the modified ConvNeXt v2 network is further processed by a layer normalization and a feed-forward layer with $K$ nodes.
The 1D downsampling convolutional layer (channel size $=K/2$ and stride $=D$) serves as the ultimate component in the amplitude/phase sub-encoder, further downsampling the output features of the feed-forward layer by a factor of $D$ and reducing its dimensionality by half to generate amplitude/phase continuous latent code $\bm{C}_A$/$\bm{C}_P\in \mathbb{R}^{F_c\times (K/2)}$.
Therefore, we have $F_c=F/D$.

\begin{figure}
    \centering
    \includegraphics[width=\linewidth]{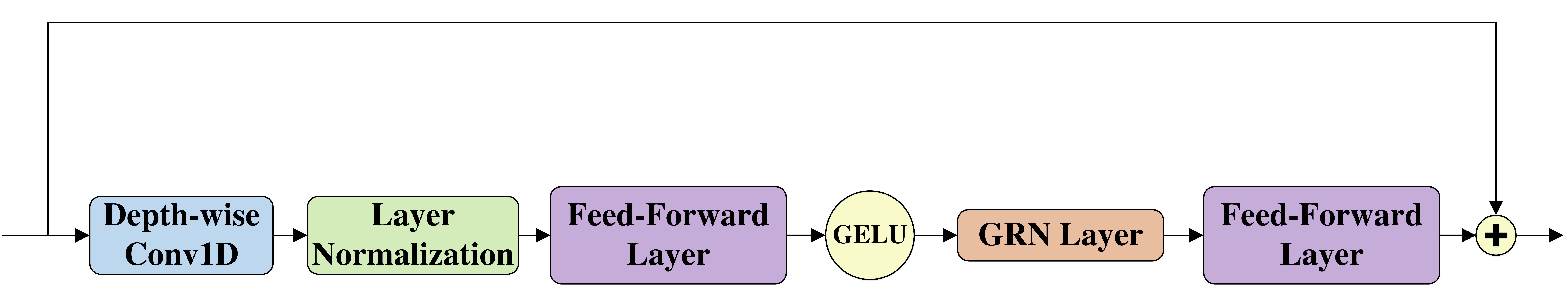}
    \caption{Details of the modified ConvNeXt v2 blcok. Here, \emph{Conv1D}, \emph{GELU} and \emph{GRN} represent the 1D convolutional layer, Gaussian error linear unit and global response normalization, respectively.
    }
    \label{fig: ConVNeXtv2}
\end{figure}

The modified ConvNeXt v2 network, which is constructed by cascading 8 identical modified ConvNeXt v2 blocks, serves as the backbone of both encoder and decoder.
The modified ConvNeXt v2 block has been adapted from the ConvNeXt v2 block originally designed for image processing \cite{woo2023convnext}.
The primary modification involves replacing 2D convolutions with 1D ones, thereby tailoring the block to better suit the processing of audio signals.
As shown in Fig. \ref{fig: ConVNeXtv2}, in each modified ConvNeXt v2 block, the input sequentially passes through a 1D depth-wise convolutional layer (channel size $=K$), a layer normalization, a feed-forward layer with $K_H$ nodes that projects features into a higher dimensionality (i.e., $K_H>K$), a Gaussian error linear unit (GELU) activation \cite{hendrycks2016gaussian}, a global response normalization (GRN) layer \cite{woo2023convnext}, an another feed-forward layer with $K$ nodes that projects features back to the original dimensionality, and finally superimposes with the input (i.e., residual connections) to obtain the output.

To aggregate both the amplitude and phase information, we concatenate the amplitude code and phase code along the dimension axis to obtain a fused latent code $\left[\bm{C}_A,\bm{C}_P\right]\in \mathbb{R}^{F_c\times K}$.
Then, a dimensionality-reduction 1D convolutional layer (channel size $=N_c$) is used to significantly reduce the dimension of this fused code, resulting in a low-dimensional fused continuous latent code $\bm{C}\in \mathbb{R}^{F_c\times N_c}$ which combines both the amplitude and phase information.
The reason for reducing the dimensionality of the continuous latent code $\bm{C}$ is to concurrently decrease the dimensionality of codebooks during subsequent quantization process, facilitating the storage and transmission of codebooks.
The frame rate of $\bm{C}$ is $f_s/w_s/D$, which is one-$D$ of the frame rate of the amplitude and phase spectra.

Therefore, the functionality of the encoder can be expressed by the following formula:
\begin{align}
\label{equ: Encoder}
\bm{C}=Encoder(\bm{A},\bm{P}).
\end{align}

\subsubsection{Quantizer}
\label{subsubsec: Quantizer}

As illustrated in Fig. \ref{fig: APCodec}, the quantizer discretizes the continuous latent code $\bm{C}\in \mathbb{R}^{F_c\times N_c}$ and generates the quantized latent code $\hat{\bm{C}}\in \mathbb{R}^{F_c\times N_c}$ based on trainable codebooks.
RVQ strategy is utilized in the quantizer to lower the bitrate.
The quantizer consists of $Q$ vector quantizers (VQs), each of which has a trainable codebook $\bm{B}^q\in \mathbb{R}^{N_c\times M},q=1,\dots,Q$, where $M$ represents the number of vectors.
The quantization process is as follows.
For the first VQ, the input is the continuous latent code $\bm{C}$ and we let $\bm{L}^1=\bm{C}$.
Taking the $i$-th ($i=1,2,\dots,F_c$) frame of $\bm{L}^1$, denoted as $\bm{l}_i^1\in \mathbb{R}^{N_c}$, as an example, we first calculate the Euclidean distance between $\bm{l}_i^1$ and each vector in $\bm{B}^1$, then choose the vector with the smallest distance as the quantized code $\hat{\bm{l}}_i^1\in \mathbb{R}^{N_c}$, and save its index in $\bm{B}^1$ as $m_i^1\in\{1,2,\dots,M \}$.
Therefore, for all frames, the quantized code and index can be represented as $\hat{\bm{L}}^1=[\hat{\bm{l}}_1^1,\dots,\hat{\bm{l}}_i^1,\dots,\hat{\bm{l}}_{F_c}^1 ]^\top\in \mathbb{R}^{F_c\times N_c}$ and $\bm{m}^1=[ m_1^1,\dots,m_i^1,\dots,m_{F_c}^1 ]^\top\in \mathbb{R}^{F_c}$, respectively.
Finally, the quantization residual $\bm{L}^2=\bm{L}^1-\hat{\bm{L}}^1$ is computed as the input for the next VQ.
Repeat the above process sequentially until the final VQ's operation is completed.
The quantizer eventually generates the quantized latent code as the sum of the outputs from each VQ, i.e., $\hat{\bm{C}}=\sum_{q=1}^Q \hat{\bm{L}}^q$.
The VQ index vectors (i.e., discrete tokens) $\bm{m}^1,\bm{m}^2,\dots,\bm{m}^Q$ are represented in binary.
Therefore, the bitrate measured in kbps of the APCodec can be calculated as follows:
\begin{align}
\label{equ: Bitrate}
Bitrate=\dfrac{1}{1000}\cdot\dfrac{f_s}{w_s \cdot D}\cdot Q \cdot \log_2M.
\end{align}

The functionality of the quantizer can be expressed by the following formula:
\begin{align}
\label{equ: Quantizer}
\hat{\bm{C}},\bm{m}^1,\bm{m}^2,\dots,\bm{m}^Q=Quantizer(\bm{C}|\bm{B}^1,\bm{B}^2,\dots,\bm{B}^Q).
\end{align}
For applications such as audio communication, discrete tokens (binary form) $\bm{m}^1,\bm{m}^2,\dots,\bm{m}^Q$ are sent from the transmitter to the receiver.
The receiver then transforms the discrete tokens into quantized codes based on the codebooks and proceeds with the subsequent decoding process.
For downstream tasks such as TTS, discrete tokens are used as intermediate representations to bridge text and speech.

\subsubsection{Decoder}
\label{subsubsec: Decoder}

As illustrated in Fig. \ref{fig: APCodec}, the decoder decodes the log amplitude spectrum $\hat{\bm{A}}\in \mathbb{R}^{F\times N}$ and phase spectrum $\hat{\bm{P}}\in \mathbb{R}^{F\times N}$ in parallel from the input quantized latent code $\hat{\bm{C}}\in \mathbb{R}^{F_c\times N_c}$, and finally reconstructs the decoded waveform $\hat{\bm{x}}\in \mathbb{R}^T$ through ISTFT.
The structure of the decoder is roughly symmetrical to that of the encoder.
The parallel amplitude and phase sub-decoders are the primary components of the decoder.
The quantized latent code $\hat{\bm{C}}$ is first dimensionally restored through a 1D dimensionality-augmentation convolutional layer (channel size $=K/2$) to be used as input for both the amplitude and phase sub-decoders.
In the amplitude sub-decoder, the input is first upsampled $D$ times through a 1D deconvolutional layer (channel size $=K$ and stride $=D$), a layer normalization and then processed by a modified ConvNeXt v2 network along with a layer normalization and a feed-forward layer with $S$ nodes.
Finally, a 1D output convolutional layer (channel size $=N$) is adopted to predict the decoded log amplitude spectrum $\hat{\bm{A}}$.
The sole distinction between the phase sub-decoder and the amplitude sub-decoder lies in the utilization of a phase parallel estimation architecture proposed in our previous publication \cite{ai2023neural} at the output end of the phase sub-decoder.
The parallel estimation architecture ensures the direct prediction of wrapped phase spectra, consisting of two identical parallel 1D convolutional layers (channel size $=N$) and a phase calculation formula $\bm{\Phi}$.
Assume the outputs of the two parallel layers are $\hat{\bm{R}}\in \mathbb{R}^{F\times N}$ and $\hat{\bm{I}}\in \mathbb{R}^{F\times N}$, respectively, then the phase spectrum is calculated by $\hat{\bm{P}}=\bm{\Phi}(\hat{\bm{R}},\hat{\bm{I}})$.
Function $\bm{\Phi}$ is calculated element-wise.
For $\forall R\in \mathbb{R}$ and $I\in \mathbb{R}$, we have
\begin{align}
\label{equ: Phase calculation}
\bm{\Phi}(R,I)=\arctan\left(\dfrac{I}{R}\right)-\dfrac{\pi}{2}\cdot Sgn^*(I)\cdot\left[Sgn^*(R)-1\right],
\end{align}
and $\bm{\Phi}(0,0)=0$.
When $z\ge 0$, $Sgn^*(z)=1$; or, $Sgn^*(z)=-1$.

Therefore, the functionality of the decoder and the process of decoded waveform reconstruction can be expressed by the following formula:
\begin{align}
\label{equ: Decoder}
\hat{\bm{A}},\hat{\bm{P}}=Decoder(\hat{\bm{C}}),
\end{align}
\begin{align}
\label{equ: S}
\hat{\bm{S}}=\exp{(\hat{\bm{A}})}\cdot \exp{(j\hat{\bm{P}})},
\end{align}
\begin{align}
\label{equ: ISTFT}
\hat{\bm{x}}=ISTFT(\hat{\bm{S}}),
\end{align}
where $\hat{\bm{S}}\in \mathbb{C}^{F\times N}$ is the decoded short-time complex spectrum.

\begin{figure*}
    \centering
    \includegraphics[width=\textwidth]{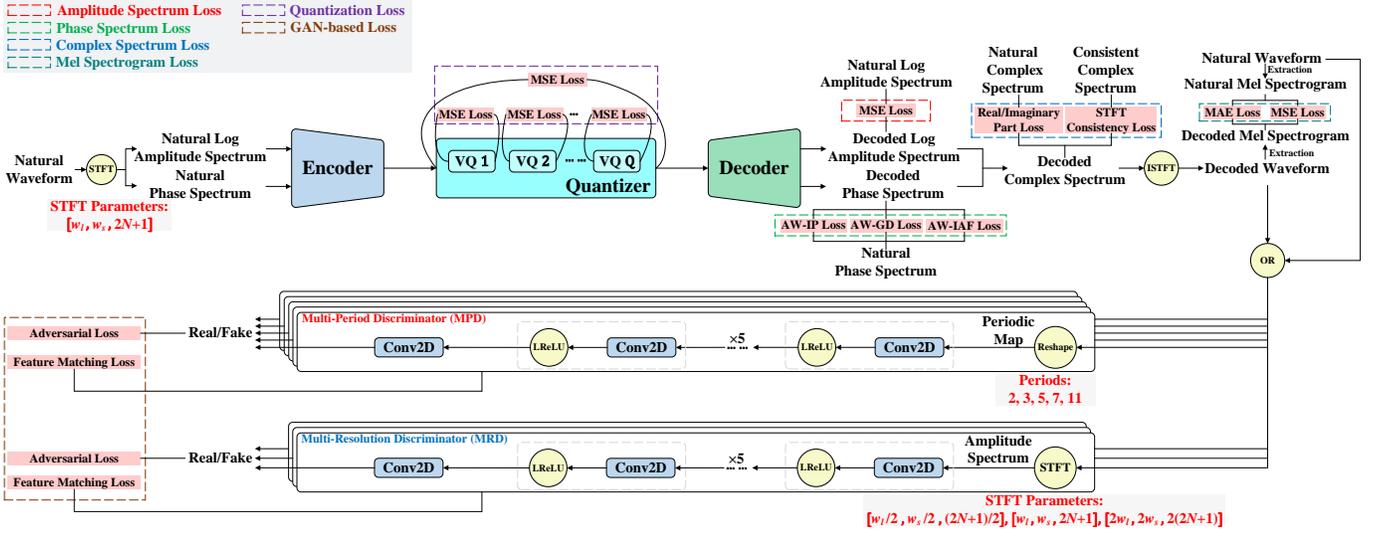}
    \caption{Details of the training losses of the proposed APCodec. Here, \emph{VQ}, \emph{Conv2D} and \emph{LReLU} represent the vector quantizer, 2D convolutional layer and leaky rectified linear unit, respectively. \emph{MSE}, \emph{MAE}, \emph{AW-IP}, \emph{AW-GD} and \emph{AW-IAF} represent mean square error, mean absolute error, anti-wrapping instantaneous phase, anti-wrapping group delay and anti-wrapping instantaneous angular frequency, respectively. \emph{STFT} and \emph{ISTFT} represent the short-time Fourier transform and inverse short-time Fourier transform, respectively. The structure of the encoder, quantizer, and decoder is simplified.
    }
    \label{fig: Loss}
\end{figure*}

\subsection{Training Criteria}
\label{subsec: Training Criteria}

A comprehensive combination of spectral-level loss, quantization loss and GAN-based loss are employed to jointly train the encoder, quantizer, and decoder in the APCodec.
These loss functions ensure the faithful reproduction of decoded audio in a comprehensive manner, highlighting how APCodec has assimilated the advantages of waveform codecs.
These losses are visualized in Fig. \ref{fig: Loss}.

\subsubsection{Spectral-level Loss}
\label{subsubsec: Spectral-level Loss}

The spectral-level loss is defined on the amplitude spectrum, phase spectrum, short-time complex spectrum and mel spectrogram, respectively, inspired by our previous publications \cite{ai2023neural,ai2023apnet}.

The loss defined on the amplitude spectrum $\mathcal L_{A}$ is the mean square error (MSE) between the decoded log amplitude spectrum $\hat{\bm{A}}\in \mathbb{R}^{F\times N}$ and the natural one $\bm{A}\in \mathbb{R}^{F\times N}$, i.e.,
\begin{align}
\label{equ: Amplitude Loss}
\mathcal L_{A}=\dfrac{1}{FN}\cdot\mathbb{E}_{\left(\hat{\bm{A}},\bm{A}\right)}\left\lVert\hat{\bm{A}}-\bm{A}\right\rVert_F^2,
\end{align}
where $\left\lVert\cdot\right\rVert_F$ denotes the Frobenius norm.

The loss defined on the phase spectrum $\mathcal L_{P}$ consists of anti-wrapping instantaneous phase (AW-IP) loss $\mathcal L_{IP}$, anti-wrapping group delay (AW-GD) loss $\mathcal L_{GD}$ and anti-wrapping instantaneous angular frequency (AW-IAF) loss $\mathcal L_{IAF}$ which are all defined between the decoded phase spectrum $\hat{\bm{P}}\in \mathbb{R}^{F\times N}$ and the natural one $\bm{P}\in \mathbb{R}^{F\times N}$.
To avoid the training error expansion issue caused by phase wrapping, we activate phase errors using an anti-wrapping function $f_{AW}(x)=\left| x-2\pi\cdot round\left( \frac{x}{2\pi} \right) \right|,x\in\mathbb R$.
The definitions of these three losses are as follows:
\begin{align}
\label{equ: AW-IP Loss}
\mathcal L_{IP}=\dfrac{1}{FN}\cdot\mathbb{E}_{\left(\hat{\bm{P}},\bm{P}\right)}\left\lVert f_{AW}\left(\hat{\bm{P}}-\bm{P}\right)\right\rVert_1,
\end{align}
\begin{align}
\label{equ: AW-GD Loss}
\mathcal L_{GD}=\dfrac{1}{FN}\cdot\mathbb{E}_{\left(\hat{\bm{P}},\bm{P}\right)}\left\lVert f_{AW}\left(\Delta_{DF}\hat{\bm{P}}-\Delta_{DF}\bm{P}\right)\right\rVert_1,
\end{align}
\begin{align}
\label{equ: AW-IAF Loss}
\mathcal L_{IAF}=\dfrac{1}{FN}\cdot\mathbb{E}_{\left(\hat{\bm{P}},\bm{P}\right)}\left\lVert f_{AW}\left(\Delta_{DT}\hat{\bm{P}}-\Delta_{DT}\bm{P}\right)\right\rVert_1,
\end{align}
where $\left\lVert\cdot\right\rVert_1$ denotes the L1 norm (entrywise form).
$\Delta_{DF}$ and $\Delta_{DT}$ represent the differential along the frequency axis and time axis, respectively.
$\mathcal L_{P}$ is the sum of the AW-IP loss, AW-GD loss and AW-IAF loss, i.e.,
\begin{align}
\label{equ: Phase Loss}
\mathcal L_{P}=\mathcal L_{IP}+\mathcal L_{GD}+\mathcal L_{IAF}.
\end{align}

Furthermore, we also establish the short-time complex spectrum loss, denoted as $\mathcal L_{S}$, to quantify the error in the decoded short-time complex spectrum $\hat{\bm{S}}\in \mathbb{C}^{F\times N}$ (i.e., Equation \ref{equ: S}).
This loss encompasses the real and imaginary part loss $\mathcal L_{RI}$, as well as a consistency loss $\mathcal L_{C}$.
$\mathcal L_{RI}$ is defined as the mean absolute error (MAE) between the real and imaginary parts of $\hat{\bm{S}}$ and the natural ones, i.e.,
\begin{align}
\label{equ: RI Loss}
\resizebox{\linewidth}{!}{
$\mathcal L_{RI}=\dfrac{1}{FN}\cdot\mathbb{E}_{\left(\hat{\bm{S}},\bm{S}\right)}\left( \left\lVert Re(\hat{\bm{S}})-Re(\bm{S})\right\rVert_1 + \left\lVert Im(\hat{\bm{S}})-Im(\bm{S})\right\rVert_1\right),$
}
\end{align}
where $\bm{S}\in \mathbb{C}^{F\times N}$ is the natural short-time complex spectrum extracted from $\bm{x}$.
$Re$ and $Im$ are the real part calculation and imaginary part calculation, respectively.
It reflects the differences between the decoded short-time complex spectrum and the natural one.
To mitigate the inconsistency issues of STFT and narrow the consistency gap between $\hat{\bm{S}}$ and the consistent short-time complex spectrum $\tilde{\bm{S}}=STFT(ISTFT(\hat{\bm{S}}))$, we define the consistency loss as follows:
\begin{align}
\label{equ: C Loss}
\resizebox{\linewidth}{!}{
$\mathcal L_{C}=\dfrac{1}{FN}\cdot\mathbb{E}_{\left(\hat{\bm{S}},\tilde{\bm{S}}\right)}\left( \left\lVert Re(\hat{\bm{S}})-Re(\tilde{\bm{S}})\right\rVert_F^2 + \left\lVert Im(\hat{\bm{S}})-Im(\tilde{\bm{S}})\right\rVert_F^2\right).$
}
\end{align}
$\mathcal L_{S}$ is a linear combination of $\mathcal L_{RI}$ and $\mathcal L_{C}$, i.e.,
\begin{align}
\label{equ: S Loss}
\mathcal L_{S}=\lambda_{RI}\mathcal L_{RI}+\mathcal L_{C},
\end{align}
where $\lambda_{RI}$ is hyperparameter.

Ultimately, we articulate the loss on the mel spectrogram as a fusion of MAE and MSE between the extracted mel spectrogram $\hat{\bm{M}}\in \mathbb{R}^{F\times N_{mel}}$ and $\bm{M}\in \mathbb{R}^{F\times N_{mel}}$ derived from $\hat{\bm{x}}$ and $\bm{x}$, respectively, i.e.,
\begin{align}
\label{equ: M Loss}
\mathcal L_{M}=\dfrac{1}{FN_{mel}}\cdot\mathbb{E}_{\left(\hat{\bm{M}},\bm{M}\right)}\left( \left\lVert \hat{\bm{M}}-\bm{M}\right\rVert_1 + \left\lVert \hat{\bm{M}}-\bm{M}\right\rVert_F^2\right),
\end{align}
where $N_{mel}$ is the dimensionality of the mel spectrogram.

Overall, the spectral-level loss is a linear combination of $\mathcal L_{A}$, $\mathcal L_{P}$, $\mathcal L_{S}$ and $\mathcal L_{M}$, i.e.,
\begin{align}
\label{equ: spectral-level Loss}
\mathcal L_{spec}=\mathcal L_{A}+\lambda_{P}\mathcal L_{P}+\lambda_{S}\mathcal L_{S}+\lambda_{M}\mathcal L_{M},
\end{align}
where $\lambda_{P}$, $\lambda_{S}$ and $\lambda_{M}$ are hyperparameters.

\subsubsection{Quantization Loss}
\label{subsubsec: Quantization Loss}

The quantization loss $\mathcal L_{Q}$ aims to reduce quantization errors, defined as the MSE between the input and output of the quantizer, as well as the MSE between the input and output of each VQ within the quantizer, i.e.,
\begin{align}
\label{equ: quantization Loss}
\resizebox{\linewidth}{!}{
$\mathcal L_{Q}=\dfrac{1}{F_cN_c}\cdot\mathbb{E}_{\left(\hat{\bm{C}},\bm{C}\right)}\left\lVert\hat{\bm{C}}-\bm{C}\right\rVert_F^2 + \dfrac{1}{F_cN_c}\cdot\sum_{q=1}^Q \mathbb{E}_{\left(\hat{\bm{L}}^q,\bm{L}^q\right)}\left\lVert\hat{\bm{L}}^q-\bm{L}^q\right\rVert_F^2.$
}
\end{align}
The quantization loss $\mathcal L_{Q}$ updates the parameters of the encoder and quantizer separately through gradient detachment operation.

\subsubsection{GAN-based Loss}
\label{subsubsec: GAN-based Loss}

For GAN-based loss, the APCodec incorporates a multi-period discriminator (MPD) \cite{kong2020hifi} to capture audio periodic patterns and a multi-resolution discriminator (MRD) \cite{jang2021univnet} to ensure the high quality of the audio spectrum across various time and frequency scales.
As shown in Fig. \ref{fig: Loss}, the MPD comprises 5 parallel sub-MPDs.
Each sub-MPD reshapes the input decoded waveform $\hat{\bm{x}}$ or natural waveform $\bm{x}$ into a 2D periodic map according to the set period.
This periodic map is subsequently processed through 5 sequential blocks, each of which consists of a 2D convolutional layer and a leaky rectified linear unit (LReLU) activation \cite{maas2013rectifier}.
Finally, the output undergoes further processing through a 2D output convolutional layer to produce a discriminative score.
The periods are respectively set to 2, 3, 5, 7, and 11.

As shown in Fig. \ref{fig: Loss}, the MRD comprises 3 parallel sub-MRDs.
Each sub-MRD extracts the amplitude spectrum from $\hat{\bm{x}}$ or $\bm{x}$ according to specified STFT parameters.
Subsequently, the amplitude spectrum undergoes processing through a network identical to that of the sub-MRD (with different convolutional layer parameters), resulting in the output of a discriminative score.
Assuming the STFT parameters for extracting the input amplitude and phase spectra for the encoder are: [frame length, frame shift, FFT point number] = [$w_l$, $w_s$, $2N+1$].
We set the STFT parameters for the three sub-MRDs as [$w_l/2$, $w_s/2$, $(2N+1)/2$], [$w_l$, $w_s$, $2N+1$] and [$2w_l$, $2w_s$, $2(2N+1)$], respectively.

The adversarial loss in the form of hinge GAN is utilized.
For a certain sub-discriminator $D^*$ in MPD and MRD, the adversarial losses for generator and discriminator are as follows:
\begin{align}
\label{equ: adv G}
\mathcal L_{adv-G}^*=\mathbb{E}_{\hat{\bm{x}}}\max \left(0,1-D^*(\hat{\bm{x}})\right),
\end{align}
\begin{align}
\label{equ: adv D}
\resizebox{\linewidth}{!}{
$\mathcal L_{adv-D}^*=\mathbb{E}_{\left(\hat{\bm{x}},\bm{x}\right)}\left[\max \left(0,1-D^*(\bm{x})\right) + \max \left(0,1+D^*(\hat{\bm{x}})\right)\right].$
}
\end{align}
Additionally, feature matching loss $\mathcal L_{FM}^*$ \cite{kumar2019melgan} is utilized, characterized by the summation of the MAE between the corresponding intermediate layer outputs of sub-discriminator $D^*$ when provided with inputs $\hat{\bm{x}}$ or $\bm{x}$.

Therefore, the GAN-based losses for generator and discriminator are respectively defined by the following expressions:
\begin{align}
\label{equ: GAN G}
\resizebox{\linewidth}{!}{
$\mathcal L_{G}=\sum_{i=1}^5 \left(\mathcal L_{adv-G}^{Pi}+\mathcal L_{FM}^{Pi} \right)+\lambda_{MRD}\sum_{j=1}^3 \left(\mathcal L_{adv-G}^{Rj}+\mathcal L_{FM}^{Rj} \right),$
}
\end{align}
\begin{align}
\label{equ: GAN D}
\resizebox{6.5cm}{!}{
$\mathcal L_{D}=\sum_{i=1}^5 \mathcal L_{adv-D}^{Pi}+\lambda_{MRD}\sum_{j=1}^3 \mathcal L_{adv-D}^{Rj},$
}
\end{align}
where the superscripts $Pi$ and $Rj$ represent $i$-th sub-MPD and $j$-th sub-MRD, respectively.
$\lambda_{MRD}$ is hyperparameter.

\subsubsection{Training Process}
\label{subsubsec: Training Process}

The final generator loss is a linear combination of the aforementioned spectral-level loss, quantization loss and GAN-based loss, i.e.,
\begin{align}
\label{equ: G_final}
\mathcal L = \lambda_{spec}\mathcal L_{spec}+\lambda_{Q}\mathcal L_{Q}+\mathcal L_{G},
\end{align}
where $\lambda_{spec}$ and $\lambda_{Q}$ are hyperparameters.
The training of the APCodec follows the standard training process of GAN, i.e., using $\mathcal L$ and $\mathcal L_{D}$ to train the generator (i.e., the encoder, quantizer and decoder) and discriminators (i.e., the MPD and MRD) alternately.

\subsection{Low-latency Implementation by Knowledge Distillation}
\label{subsec: Low-latency Implementation by Knowledge Distillation}

To attain low-latency streamable inference, we implement modifications to specific components of the APCodec, covering the following three aspects.
1) Unlike some well-known codecs like SoundStream \cite{zeghidour2021soundstream} and Encodec \cite{defossez2022high} that employ causal convolutions, the streamable APCodec replaces all non-causal convolutional layers (excluding upsampling/downsampling layers) with feed-forward layers.
This approach is conducive to reducing model size and improving generation efficiency;
2) Replacing the original non-causal upsampling deconvolutional layers with causal ones;
3) Setting the kernel size of the downsampling convolutional layer to be smaller than or equal to $2D-1$.
For the last aspect, we provide a detailed explanation as follows.
To ensure the generation of at least one frame of latent code, the minimum length of the input audio for APCodec is $w_s\cdot D$ samples (i.e., fixed latency).
Therefore, the input features of the downsampling convolutional layer have at least $D$ frames. During the convolution operation, $D-1$ zeros are padded before the features. Hence, downsampling convolutional operations can be performed without utilizing future information when the kernel size is smaller than or equal to $2D-1$.

However, with the aforementioned modifications, the streamable APCodec, compared to the original non-streamable APCodec, inevitably leads to a deterioration in decoded audio quality.
Therefore, we introduce a knowledge distillation training strategy, utilizing a well-trained non-streamable APCodec as the teacher model to guide the training of the streamable APCodec (i.e., the student model).
To establish a connection between the teacher model and the student model, we introduce a knowledge distillation loss $\mathcal L_{KD}$, defined as the MSE between the features of the two models at corresponding positions.
These positions encompass the output of all convolutional layers, feed-forward layers, modified ConvNeXt v2 blocks, and the quantizer in Fig. \ref{fig: APCodec}.
At the training stage, the streamable APCodec uses $\mathcal L+\lambda_{KD}\mathcal L_{KD}$ and $\mathcal L_{D}$ to train the generator and discriminators alternately, where $\lambda_{KD}$ is hyperparameter.
The other hyperparameters used for training the streamable APCodec, as well as the dataset and the total training steps, are entirely consistent with training the non-streamable APCodec.
Through training, the streamable APCodec aims to approach the decoded audio quality of the non-streamable APCodec while maintaining its advantage of low latency.

\section{Experiments}
\label{sec: Experiments}

\subsection{Data and Feature Configuration}
\label{subsec: Data and Feature Configuration}

A subset of the VCTK-0.92 corpus \cite{yamagishi2019cstr} which contained approximately 43 hours of 48 kHz speech recordings from 108 speakers with various accents, was adopted in our experiments.
We selected 40,936 utterances from 100 speakers as the training set.
Then we built the test set, which included 2,937 utterances from remaining 8 unseen speakers.
The original 48 kHz waveforms and downsampled waveforms at 24 kHz and 16 kHz were used for the experiments (i.e., $f_s = 48000$ or $24000$ or $16000$).
When extracting the amplitude spectra, phase spectra and mel spectrograms from natural waveforms, the window size was 320 samples (i.e., $w_l = 320$), the window shift was 40 samples (i.e., $w_s = 40$), and the FFT point number was 1024 (i.e., $N = 513$).
The dimensionality of the mel spectrograms was 80 (i.e., $N_{mel}=80$).
This configuration is applicable to waveforms at all sampling rates.

\subsection{Model Details}
\label{subsec: Model Details}

In our experiments\footnote{Source codes are available at \url{https://github.com/yangai520/APCodec}. Examples of generated audio can be found at \url{https://yangai520.github.io/APCodec}.}, we constructed non-streamable and streamable APCodec models to fairly compare with existing non-streamable and streamable codec models, respectively.
The descriptions of the non-streamable and streamable APCodec are as follows.

\begin{itemize}

\item {}{\textbf{APCodec}}: The proposed non-streamable APCodec.
In the encoder and decoder, the kernel size for all convolutional operations was set to 7.
The kernel size for two deconvolutional operations was set to 16.
The channel size $K$, $K_H$ and $N_c$ were 256, 512 and 32, respectively.
The downsampling/upsampling ratio was $D=8$.
Fig. \ref{fig: APCodec} serves as an example of a 48 kHz audio codec, showcasing the frame rates of the spectral characteristics and latent codes.
We can observe that the APCodec only requires 8$\times$ downsampling to encode a latent code with a frame rate as low as 150 Hz.
The hyperparameters for loss functions were set as $\lambda_P=\frac{20}{9}$, $\lambda_{RI}=2.25$, $\lambda_S=\frac{4}{9}$, $\lambda_M=1$, $\lambda_{spec}=45$, $\lambda_Q=7.5$, and $\lambda_{MRD}=0.1$.
The model was trained using the AdamW optimizer \cite{loshchilov2018decoupled} with $\beta_1=0.8$ and $\beta_2=0.99$ on a single NVIDIA RTX 3090 GPU.
The learning rate decay was scheduled by a 0.999 factor in every epoch with an initial learning rate of 0.0002.
The batch size was 16, and the truncated waveform length was 7960 samples for each training step.
The model was trained until 1M steps.

\item {}{\textbf{APCodec-S}}: The proposed streamable APCodec.
It was modified according to the methods outlined in Section \ref{subsec: Low-latency Implementation by Knowledge Distillation} for the \textbf{APCodec}, wherein the number of nodes in the replaced feed-forward layers remain consistent with the original convolutional layers' channel size.
The kernel size of the downsampling convolutional layers was set to 7.
It is trained with guidance from the well-trained \textbf{APCodec}.
The hyperparameter for knowledge distillation was set as $\lambda_{KD}=1$.
Other training strategies were consistent with those used in \textbf{APCodec}.

\end{itemize}

For high-sampling-rate audio coding, we compared the proposed APCodec with the following codecs:

\begin{itemize}

\item {}{\textbf{Encodec}}: The Encodec \cite{defossez2022high} audio codec.
It supports audio coding at sampling rates of 24 kHz and 48 kHz, as reported in \cite{defossez2022high}.
We reimplemented it using the open source implementation\footnote{\url{https://github.com/yangdongchao/AcademiCodec}.}.
The downsampling/upsampling ratio was 320.
It can achieve low-latency streamable inference.

\item {}{\textbf{AudioDec}}: The AudioDec \cite{wu2023audiodec} audio codec.
It is specifically designed for 48 kHz audio codec.
We reimplemented the \emph{AudioDec v1} model in \cite{wu2023audiodec} which has been confirmed to deliver the best performance using the open source implementation\footnote{\url{https://github.com/facebookresearch/AudioDec}.}.
This model integrates both the encoder and the HiFi-GAN vocoder.
Therefore, the \emph{AudioDec v1} model is not an end-to-end model.
The downsampling/upsampling ratio for the model was 320.
The AudioDec can also achieve low-latency streamable inference.

\item {}{\textbf{DAC}}: The DAC \cite{kumar2024high} audio codec.
It is designed for 44.1 kHz audio coding.
We reimplemented it using the open source implementation\footnote{\url{https://github.com/descriptinc/descript-audio-codec}.} and applied it to 48 kHz audio coding.
The downsampling/upsampling ratio for the model was 320.
However, the DAC is non-streamable, and there is no streamable implementation provided in the open-source code.

\end{itemize}

Although our proposed APCodec was initially designed for 48 kHz audio coding, to ensure fair comparisons with some low-sampling-rate audio codecs, we also conducted experiments at lower sampling rates, such as 16 kHz and 24 kHz.
In addition to \textbf{Encodec}, the low-sampling-rate audio codecs used for comparison also included the following:

\begin{itemize}

\item {}{\textbf{SoundStream}}: The SoundStream \cite{zeghidour2021soundstream} audio codec.
We reimplemented it using the open source implementation$^\text{2}$.
The downsampling/upsampling ratio was 320.
It can achieve low-latency streamable inference.

\item {}{\textbf{HiFi-Codec}}: The HiFi-Codec \cite{yang2023hifi} audio codec.
We reimplemented it using the open source implementation$^\text{2}$.
The downsampling/upsampling ratio was also 320.
However, the HiFi-Codec is non-streamable, and there is no streamable implementation provided in the open-source code.

\end{itemize}

These codecs were comparable because they all employed the similar quantization method (i.e., RVQ or related strategy).
All of the above codecs adopted 1024 vectors (i.e., $M=1024$) in the codebook of each VQ.
We conducted experiments at all three sampling rates, with two bitrates (low and high) tested at each sampling rate.
For 48 kHz audio codecs, the bitrates were set at 6 kbps and 12 kbps, respectively.
For 24 kHz audio codecs, the bitrates were set at 3 kbps and 6 kbps, respectively.
For 16 kHz audio codecs, the bitrates were set at 2 kbps and 4 kbps, respectively.
The configuration of low bitrate and high bitrate for \textbf{APCodec}, \textbf{APCodec-S}, \textbf{Encodec}, \textbf{AudioDec}, \textbf{DAC} and \textbf{SoundStream} was achieved by setting the number of VQs within the quantizer to $Q=4$ and $Q=8$, respectively.
Due to the adoption of the GRVQ quantization strategy in \textbf{HiFi-Codec}, it employed two groups of RVQ, each consisting of 2 and 4 VQs, for achieving audio coding for both low and high bitrates, respectively.

\subsection{Evaluation Metrics}
\label{subsec: Evaluation Metrics}

First, we comprehensively evaluated the performance of these compared audio codecs using multiple objective metrics.
These objective metrics were specifically designed to evaluate the amplitude spectrum quality, overall audio objective quality, intelligibility, phase spectrum quality, generation speed and model complexity, respectively.

\begin{itemize}

\item {}{\textbf{Amplitude spectrum quality}}: The commonly used log-spectral distance (LSD) and mel-cepstrum distortion (MCD) were employed to evaluate the amplitude spectrum quality between the decoded audio $\hat{\bm{x}}$ generated by a codec and the natural one $\bm{x}$.
    The LSD and MCD respectively represented the distortion of audio in the log amplitude spectral domain and the mel cepstral domain.
    A smaller result indicates less distortion.


\item {}{\textbf{Intelligibility}}: The commonly used short-time objective intelligibility (STOI) \cite{taal2010short} was used to quantify the intelligibility of $\hat{\bm{x}}$, with natural audio $\bm{x}$ as the reference.
    The STOI score ranges from 0 to 1.
    A higher STOI score indicates that the speech is more easily understandable to humans.

\item {}{\textbf{Overall audio objective quality}}: The commonly used virtual speech quality objective listener (ViSQOL)\footnote{\url{https://github.com/google/visqol}.} \cite{chinen2020visqol} tool was used to objectively assess the overall quality of the decoded audio $\hat{\bm{x}}$, with natural audio $\bm{x}$ as the reference.
    The ViSQOL outputs a mean opinion score - listening quality objective (MOS-LQO) score, where a higher score indicates better audio quality.
    The ViSQOL supports only two sampling rates: 48 kHz and 16 kHz.
    For 48 kHz ViSQOL, the MOS-LQO ranges from 1 to 4.75.
    For 16 kHz ViSQOL, the MOS-LQO ranges from 1 to 5.
It should be noted that for the assessment of audio quality at a 24 kHz sampling rate, we upsampled both the decoded audio and reference audio to 48 kHz, and then calculated MOS-LQO using ViSQOL's 48 kHz mode.

    \item {}{\textbf{Phase spectrum quality}}: One of the highlights of the proposed APCodec lies in the phase modeling.
     To validate the effectiveness of phase modeling, the anti-wrapping phase distance (AWPD) proposed in our previous work \cite{lu2024towards}, was employed to evaluate the phase spectrum quality between $\hat{\bm{x}}$ and $\bm{x}$.
    Similar to the mentioned phase loss in Section \ref{subsubsec: Spectral-level Loss}, the AWPD was also computed separately for instantaneous phase, group delay, and instantaneous angular frequency (denoted as $\text{AWPD}_{\text{IP}}$, $\text{AWPD}_{\text{GD}}$ and $\text{AWPD}_{\text{IAF}}$, respectively).
    After the activation of phase errors using the anti-wrapping function $f_{AW}$, the AWPD is calculated in a manner akin to LSD, allowing it to accurately depict the actual phase distortion.
    A smaller result indicates less distortion.

\item {}{\textbf{Generation speed}}: The real-time factor (RTF), which is defined as the ratio between the time consumed to generate audio waveforms and the duration of the generated audio waveforms, was utilized to evaluate the generation speed of a codec.
    In our implementation, the RTF value was calculated as the ratio between the time consumed to generate all test sentences using a single NVIDIA RTX 3090 GPU or a single Intel Xeon E5-2680 CPU core and the total duration of the test set.
    A lower RTF indicates a faster generation speed.

\item {}{\textbf{Model complexity}}: The model size (excluding the discriminators) is used to measure the complexity of the codec model.
For the application of audio codecs on certain embedded devices, a lightweight model is crucial.

\end{itemize}

Furthermore, to assess human perception of the decoded audio quality, we also conducted subjective experiments.
Since the focus of this paper is on high-sampling-rate audio coding, subjective experiments were conducted only on the 48 kHz sampling rate configuration.
We conducted a MUSHRA (MUltiple Stimuli with Hidden Reference and Anchor) test \cite{recommendation2001method} on the crowdsourcing platform Amazon Mechanical Turk\footnote{\url{https://www.mturk.com}.} to evaluate the quality of the 48 kHz audio decoded by \textbf{APCodec}, \textbf{APCodec-S}, \textbf{Encodec}, \textbf{AudioDec} and \textbf{DAC} at 6 kbps and 12 kbps on the test set of the VCTK dataset.
20 test utterances decoded by each experimental model were evaluated by about 40 English native listeners.
Listeners were asked to give a score between 0 and 100 to each test sample (the reference natural audio tracks had a maximum score of 100).

\begin{table*}
\centering
    \caption{Objective experimental results for compared codecs on the test set of the VCTK dataset at three sampling rates and two bitrates. The \textbf{bold} and \underline{underline} numbers indicate the optimal and sub-optimal results, respectively.}
    \resizebox{\textwidth}{!}{
    \begin{tabular}{c c c c | c c | c | c | c c c}
        \hline
        \hline
        \multirow{2}{*}{Codec} & Low-latency & \multirow{2}{*}{Sampling rate} & \multirow{2}{*}{Bitrate} & LSD & MCD & \multirow{2}{*}{STOI$\uparrow$} & \multirow{2}{*}{ViSQOL$\uparrow$} & $\text{AWPD}_{\text{IP}}$ & $\text{AWPD}_{\text{GD}}$ & $\text{AWPD}_{\text{IAF}}$  \\
         & (Streamable) & & & (dB)$\downarrow$ & (dB)$\downarrow$ &  &  & (rad)$\downarrow$ & (s)$\downarrow$ & (rad/s)$\downarrow$  \\
         \hline
         \hline
         \textbf{APCodec} & No & \multirow{5}{*}{48 kHz} & \multirow{5}{*}{6 kbps} & \textbf{0.818} & \textbf{1.60} & \underline{0.875} & \textbf{4.07} & \textbf{1.68} & \textbf{1.40} & \textbf{1.44}\\
         \textbf{APCodec-S} & Yes & & & \underline{0.835} & \underline{1.67} & 0.865 & 3.93 & \underline{1.74} & \underline{1.41} & \underline{1.45}\\
         \textbf{Encodec} \cite{defossez2022high} & Yes & & & 1.04 & 2.61 & 0.793 & 3.31 & 1.80 & 1.43 & 1.47 \\
         \textbf{AudioDec} \cite{wu2023audiodec} & Yes & & & 0.847 & 2.85 & 0.804 & \underline{3.98} & 1.81 & 1.44 & 1.46 \\
         \textbf{DAC} \cite{kumar2024high} & No & & & 0.841 & 1.87 & \textbf{0.906} & 3.81 & 1.78 & \textbf{1.40} & \textbf{1.44} \\
         \hline
         \textbf{APCodec} & No & \multirow{5}{*}{48 kHz} & \multirow{5}{*}{12 kbps} & \textbf{0.796} & \textbf{1.33} & 0.901 & \textbf{4.26} & \textbf{1.60} & \textbf{1.38} & \textbf{1.42} \\
         \textbf{APCodec-S} & Yes & & & 0.822 & \underline{1.42} & \underline{0.903} & 4.13 & \underline{1.70} & \underline{1.40} & \underline{1.44}\\
         \textbf{Encodec} \cite{defossez2022high} & Yes & & & 0.885 & 2.17 & 0.860 & 3.51 & 1.79 & 1.42 & 1.45 \\
         \textbf{AudioDec} \cite{wu2023audiodec} & Yes & & & 0.831 & 2.31 & 0.825 & \underline{4.14} & 1.81 & 1.44 & 1.46 \\
         \textbf{DAC} \cite{kumar2024high} & No & & & \underline{0.815} & 1.76 & \textbf{0.954} & 4.06 & 1.75 & \textbf{1.38} & \textbf{1.42} \\
         \hline
         \hline
         \textbf{APCodec} & No &  \multirow{5}{*}{24 kHz} & \multirow{5}{*}{3 kbps}  & \textbf{0.839} & 2.31 & \underline{0.856} & \underline{4.08} & \textbf{1.66} & \textbf{1.36} & \textbf{1.42} \\
         \textbf{APCodec-S} & Yes &  &  & 0.864 & \underline{2.18} & 0.838 & \textbf{4.11} & \underline{1.78} & \underline{1.38} & \underline{1.44} \\
         \textbf{Encodec} \cite{defossez2022high} & Yes & &  & 0.958 & 2.74 & 0.817 & 3.82 & 1.79 & 1.39 & \underline{1.44}\\
         \textbf{SoundStream} \cite{zeghidour2021soundstream} & Yes &  &  & 0.977 & 3.03 & 0.804 & 3.79 & 1.79 & 1.40 & \underline{1.44}\\
         \textbf{HiFi-Codec} \cite{yang2023hifi} & No & &  & \underline{0.849} & \textbf{2.10} & \textbf{0.875} & 4.05 & 1.79 & \textbf{1.36} & \underline{1.44} \\
         \hline
         \textbf{APCodec} & No &  \multirow{5}{*}{24 kHz} & \multirow{5}{*}{6 kbps}  & \underline{0.815} & 2.02 & 0.877 & \underline{4.28} & \textbf{1.60} & \textbf{1.34} & \textbf{1.41}\\
         \textbf{APCodec-S} & Yes &  &  & \textbf{0.812} & \textbf{1.64} & \underline{0.889} & \textbf{4.35} & \underline{1.66} & \underline{1.35} & \underline{1.42}\\
         \textbf{Encodec} \cite{defossez2022high} & Yes &  &  & 0.933 & 2.53 & 0.836 & 3.81 & 1.78 & 1.38 & 1.44 \\
         \textbf{SoundStream} \cite{zeghidour2021soundstream} & Yes &  &  & 0.944 & 2.70 & 0.832 & 3.90 & 1.78 & 1.38 & 1.44 \\
         \textbf{HiFi-Codec} \cite{yang2023hifi} & No &  &  & 0.850 & \underline{1.83} & \textbf{0.910} & 4.13 & 1.77 & \underline{1.35} & 1.43 \\
         \hline
         \hline
         \textbf{APCodec} & No &  \multirow{5}{*}{16 kHz} & \multirow{5}{*}{2 kbps} & \textbf{0.834} & \textbf{2.48} & \textbf{0.852} & \textbf{4.09} & \textbf{1.68} & \textbf{1.33} & \textbf{1.41} \\
         \textbf{APCodec-S} & Yes &  &  & \underline{0.856} & 2.56 & \underline{0.851} & \underline{4.05} & \underline{1.73} & \textbf{1.33} & \underline{1.42} \\
         \textbf{Encodec} \cite{defossez2022high} & Yes &  &  & 0.939 & 2.98 & 0.810 & 3.70 & 1.78 & 1.36 & 1.43 \\
         \textbf{SoundStream} \cite{zeghidour2021soundstream} & Yes &  &  & 0.965 & 3.11 & 0.804 & 3.62  & 1.78 & 1.36 & 1.44\\
         \textbf{HiFi-Codec} \cite{yang2023hifi} & No &  &  & 0.910 & \underline{2.49} & 0.832 & 3.84 & 1.79 & \underline{1.35} & 1.43 \\
         \hline
         \textbf{APCodec} & No &  \multirow{5}{*}{16 kHz} & \multirow{5}{*}{4 kbps}  & \textbf{0.792} & \underline{2.12} & \textbf{0.885} & \underline{4.32} & \textbf{1.56} & \textbf{1.29} & \textbf{1.38} \\
         \textbf{APCodec-S} & Yes &  &  & \underline{0.810} & \textbf{1.88} & \underline{0.881} & \textbf{4.35} & \underline{1.66} & \underline{1.32} & \underline{1.41} \\
         \textbf{Encodec} \cite{defossez2022high} & Yes &  &  & 0.928 & 2.78 & 0.823 & 3.77 & 1.77 & 1.35 & 1.43 \\
         \textbf{SoundStream} \cite{zeghidour2021soundstream} & Yes &  &  & 0.938 & 2.76 & 0.837 & 3.83 & 1.76 & 1.35 & 1.43 \\
         \textbf{HiFi-Codec} \cite{yang2023hifi} & No &  &  & 0.875 & 2.14 & 0.869 & 4.10 & 1.77 & 1.33 & 1.42 \\

        \hline
        \hline
    \end{tabular}}
\label{tab_objective_subjective_48}
\end{table*}

\subsection{Primary Experimental Results}
\label{subsec: Primary Experimental Results}

The primary experiments aim to compare the performance differences between our proposed APCodec and other neural codecs.
The APCodec is designed for high sampling rates and low bitrates, thus we focus our analysis on the audio codec results at a 48 kHz sampling rate.
The experimental results at 48 kHz are depicted in Table \ref{tab_objective_subjective_48} and Table \ref{tab_objective_subjective_RTF}.
It can be observed that at a sampling rate of 48 kHz and a bitrate of 6 kbps (low bitrate), without considering latency, our proposed APCodec achieved state-of-the-art (SOTA) performance across various metrics.
Surprisingly, the ViSQOL score of the \textbf{APCodec} reached 4.07.

Specifically, we first compared the proposed \textbf{APCodec} with \textbf{DAC} because both of them were non-streamable.
As shown in Table \ref{tab_objective_subjective_48}, at a sampling rate of 48 kHz and a bitrate of 6 kbps, the proposed \textbf{APCodec} outperformed the \textbf{DAC} significantly for most metrics, except for the STOI metric.
From the perspective of LSD and MCD, the \textbf{APCodec} exhibited higher quality in decoded audio amplitude spectrum, highlighting its advantage in explicitly modeling amplitude spectra.
Similarly, according to the results of the AWPD metrics, it is evident that explicit modeling of phase spectra in the \textbf{APCodec} contributed to improving the precision of decoded phases.
However, among the three specific metrics of AWPD, the difference reflected by $\text{AWPD}_{\text{IP}}$ was more pronounced.
The $\text{AWPD}_{\text{GD}}$ values for all codecs at 48 kHz and 6 kbps in Table \ref{tab_objective_subjective_48} were concentrated in the range of 1.40 to 1.44, while the $\text{AWPD}_{\text{IAF}}$ values were concentrated in the range of 1.44 to 1.47.
The differences in $\text{AWPD}_{\text{GD}}$ and $\text{AWPD}_{\text{IAF}}$ metrics between different codecs were minor.
Additionally, we also observed that, apart from the proposed \textbf{APCodec} and \textbf{APCodec-S}, the $\text{AWPD}_{\text{IP}}$ values for other baseline codecs were all around 1.80.
The shared characteristic among these codecs is the absence of explicit phase modeling.
Consequently, we hypothesize that their $\text{AWPD}_{\text{IP}}$ values may reflect a consistent initial phase error.
The \textbf{APCodec} we proposed achieved a reduction of approximately 0.12 in the $\text{AWPD}_{\text{IP}}$ value through explicit prediction and optimization of the phase.
The aforementioned findings suggested a similarity in the phase spectrum continuity of the decoded audio across these codecs according to the results of $\text{AWPD}_{\text{GD}}$ and $\text{AWPD}_{\text{IAF}}$.
Nevertheless, the \textbf{APCodec} standed out by producing audio with instantaneous phase values that closely aligned with the natural phase, showcasing superior quality in decoded phase.
Although the \textbf{APCodec} lagged behind the \textbf{DAC} in intelligibility (i.e., STOI), it benefited from the aforementioned advantages, placing it in a leading position in terms of overall audio objective quality (i.e., ViSQOL).
In terms of generation efficiency, as shown in Table \ref{tab_objective_subjective_RTF}, whether on GPU or CPU, the \textbf{APCodec} exhibited faster generation speed than the \textbf{DAC}.
This advantage was more pronounced when running on CPU.
The generation speed of \textbf{APCodec} on the CPU was approximately 14 times faster than that of the \textbf{DAC}, and the \textbf{DAC} was unable to achieve real-time generation on CPU.
This phenomenon indicated that without the parallel acceleration of GPU, utilizing spectra as coding objects can significantly enhance generation efficiency compared to the direct encoding and decoding of waveforms.
Especially for high-sampling-rate audio codecs, the spectrum-based approach was more suitable due to the larger number of waveform samples.
Furthermore, the \textbf{APCodec} was a lightweight model, with a model size of only 23.2\% compared to that of the the \textbf{DAC}.
Although the \textbf{APCodec} and \textbf{DAC} had similar subjective perceptual quality, according to the MUSHRA scores in Table \ref{tab_objective_subjective_RTF}, overall, the \textbf{APCodec} performed the best. This is because it demonstrated significantly faster generation speed, a lighter model, and superior performance on most objective metrics.

\begin{table*}
\centering
    \caption{Objective and subjective experimental results for compared codecs on the test set of the VCTK dataset at 48 kHz sampling rate and two bitrates. Here, ``$a\times$" represents $a\times$ real time. The \textbf{bold} and \underline{underline} numbers indicate the optimal and sub-optimal results, respectively.}
    \begin{tabular}{c c | c c | c | c}
        \hline
        \hline
         Codec & Bitrate & RTF (GPU)$\downarrow$ & RTF (CPU)$\downarrow$ & Model size$\downarrow$ & MUSHRA score \\
         \hline
         \hline
         \textbf{APCodec} & \multirow{5}{*}{6 kbps} & \underline{0.0112 (89.3$\times$)} & \underline{0.173 (5.78$\times$)} & \underline{65.4M} & \textbf{88.48$\pm$0.87}\\
         \textbf{APCodec-S} &  &  \textbf{0.0109 (91.7$\times$)} & \textbf{0.112 (8.93$\times$)} & \textbf{46.8M}  & 88.07$\pm$0.89\\
         \textbf{Encodec} \cite{defossez2022high} &  & 0.0149 (67.1$\times$) & 0.232 (4.31$\times$) & 83.2M & 85.91$\pm$1.25\\
         \textbf{AudioDec} \cite{wu2023audiodec} &  & 0.0132 (75.8$\times$) & 0.771 (1.30$\times$) & 108M & \underline{88.34$\pm$0.89}\\
         \textbf{DAC} \cite{kumar2024high} &  & 0.0195 (51.3$\times$) & 2.47 (0.405$\times$) & 282M & 88.28$\pm$0.92\\
         \hline
         \textbf{APCodec} & \multirow{5}{*}{12 kbps} & \underline{0.0120 (83.3$\times$)} & \underline{0.181 (5.52$\times$)} & \underline{66.0M} & \underline{90.68$\pm$0.86}\\
         \textbf{APCodec-S} &  &  \textbf{0.0119 (84.0$\times$)} & \textbf{0.116 (8.62$\times$)} & \textbf{47.4M} & 90.16$\pm$0.85\\
         \textbf{Encodec} \cite{defossez2022high}&  & 0.0157 (63.7$\times$) & 0.238 (4.20$\times$) & 99.2M & 88.09$\pm$1.21\\
         \textbf{AudioDec} \cite{wu2023audiodec}&  &  0.0135 (74.1$\times$) & 0.780 (1.28$\times$) & 110M & 89.66$\pm$0.93\\
         \textbf{DAC} \cite{kumar2024high}&  & 0.0216 (46.3$\times$) & 2.68 (0.373$\times$) & 283M & \textbf{90.78$\pm$0.91}\\

        \hline
        \hline
    \end{tabular}
\label{tab_objective_subjective_RTF}
\end{table*}

Then, we compared the \textbf{APCodec-S} with other streamable codecs, i.e., \textbf{Encodec} and \textbf{AudioDec} at 48 kHz and 6 kbps.
As mentioned in Section \ref{subsec: Low-latency Implementation by Knowledge Distillation}, these codecs all had an unavoidable fixed latency.
According to the current experimental setup, the latency for \textbf{APCodec-S}, \textbf{Encodec} and \textbf{AudioDec} was all approximately 6.67 ms (i.e., 320 samples) for 48 kHz audio, hence their comparison is fair.
As shown in Table \ref{tab_objective_subjective_48} and \ref{tab_objective_subjective_RTF}, the proposed \textbf{APCodec-S} significantly outperformed the baseline \textbf{Encodec} on all objective and subjective metrics.
However, we found that the \textbf{AudioDec}, which is a combination of encoder and vocoder, served as a robust baseline, with an objective ViSQOL score 0.05 higher than the \textbf{APCodec-S} and a similar subjective MUSHRA score with the \textbf{APCodec-S}.
Yet, it lagged behind the \textbf{APCodec-S} in all other metrics.
As illustrated in Table \ref{tab_objective_subjective_RTF}, the generation speed of the \textbf{Encodec} was relatively fast, slightly trailing behind the \textbf{APCodec-S}, but its model size was 1.78 times that of the \textbf{APCodec-S}.
The generation speed of the \textbf{AudioDec} on CPU was relatively slow, just reaching the real-time standard.
This may be attributed to the introduction of the HiFi-GAN vocoder.
The introduction of the vocoder also resulted in a large model size, approximately 2.3 times that of the \textbf{APCodec-S}.
Furthermore, the two-stage training paradigm of the AudioDec also led to operational complexity, in contrast to our proposed end-to-end APCodec.

By comparing the \textbf{APCodec} and \textbf{APCodec-S} at 48 kHz and 6 kbps in Table \ref{tab_objective_subjective_48}, the overall objective performance of the streamable model decreased compared to non-streamable one.
This is reasonable, as the streamable model did not leverage future information.
The \textbf{APCodec} can be considered as an upper-bound model for the \textbf{APCodec-S}.
Nevertheless, the \textbf{APCodec-S} still outperformed numerous streamable baselines.
It's worth mentioning that the \textbf{APCodec-S} had an increase of 0.06 in $\text{AWPD}_{\text{IP}}$ metric compared to the \textbf{APCodec}.
Although this difference was small, during the training process, we clearly observed that the AW-IP loss reduction was challenging for the \textbf{APCodec-S}.
This also reflects that in our proposed model, the convergence status of the AW-IP loss can be used to preliminarily estimate the quality of the decoded audio, which is helpful for model selection.
Although the introduction of low-latency implementation led to a significant deterioration in objective metrics, according to the MUSHRA scores in Table \ref{tab_objective_subjective_48}, the subjective perceptual quality only slightly declined.
Furthermore, compared to the \textbf{APCodec}, the \textbf{APCodec-S} showed improved efficiency and a further reduction in model size, as shown in Table \ref{tab_objective_subjective_RTF}.
This is because, in the process of transforming the non-streamable model into a streamable one, we chose to replace non-causal convolutions with feed-forward layers instead of causal convolutions used in Encodec and AudioDec.
This reduced the model complexity and further improved the generation speed.

To further assess the performance of the codecs at different bitrates, we conducted experiments on these comparative codecs at 48 kHz and 12 kbps.
The results are also presented in Tables \ref{tab_objective_subjective_48} and \ref{tab_objective_subjective_RTF}.
For the same codec, there was a noticeable improvement in both objective and subjective aspects at 12 kbps compared to those at 6 kbps.
Simultaneously, the former experienced a decrease in generation speed, coupled with an increase in model complexity.
This is reasonable, as increasing the number of VQs can reduce the quantization error and increase trainable parameters.
The comparison results for different codecs at 12 kbps were essentially consistent with those at 6 kbps.
Notably, the ViSQOL score for the \textbf{APCodec} at 12 kbps reached an impressive 4.26 (the maximum score is 4.75).
However, the performance of other codecs has also significantly improved.
For instance, the \textbf{DAC} achieved remarkably high intelligibility for audio decoded at 12 kbps, as indicated by the STOI results, despite it was inferior to or comparable with our proposed \textbf{APCodec} in other metrics.
In addition, the \textbf{AudioDec} also demonstrated a noticeable performance improvement at 12 kbps.
This result aligns with expectations, as the AudioDec \cite{wu2023audiodec} was originally designed as a 48 kHz audio codec for 12.8 kbps.
Fortunately, the proposed \textbf{APCodec-S} still maintained its overall superiority over \textbf{AudioDec} at 12 kbps, with the difference in ViSQOL metrics being only 0.01.
However, apart from phase metrics, the difference between APCodec and other codecs at 12 kbps was smaller compared to the difference at 6 kbps across other metrics.
This observation underscored the suitability of the proposed APCodec for encoding and decoding at low bitrates, showcasing enhanced audio compression capabilities.

Due to their original design as low sampling rate for some well-known audio codecs, e.g., SoundStream, Encodec and HiFi-Codec, we also conducted comparative experiments at sampling rates of 16 kHz and 24 kHz.
The objective experimental results are shown in Table \ref{tab_objective_subjective_48}.
It can be observed that both the streamable \textbf{Encodec} and \textbf{SoundStream} exhibited significant gaps compared to our proposed streamale \textbf{APCodec-S} at these two sampling rates.
For comparisons between non-streamable codecs, at 16 kHz, the \textbf{APCodec} surpassed the \textbf{HiFi-Codec} across all metrics.
However, at 24 kHz, the \textbf{APCodec} did not perform as well as the \textbf{HiFi-Codec} in terms of MCD and STOI.
This may be attributed to the fact that the HiFi-Codec originally excelled at a 24 kHz sampling rate \cite{yang2023hifi}.
Interestingly, when comparing the \textbf{APCodec} and \textbf{APCodec-S} at low sampling rates and high bitrates, the \textbf{APCodec-S} even outperformed \textbf{APCodec} in terms of MCD and ViSQOL metrics, which suggested an improvement in perceptual quality in the mel scale.
This means that the proposed low-latency implementation is more effective for low-sampling-rate APCodec, because the low-latency implementation under high sampling rate conditions obviously reduced the ViSQOL score as shown in Table \ref{tab_objective_subjective_48}.
The above results indicated that while our proposed APCodec exhibited a more pronounced advantage at 48 kHz, applying it at lower sampling rates also yielded good performance.

Based on the above experimental results, we can conclude that APCodec, by leveraging the advantages of parametric codec and waveform codec, is better suited for audio coding at both high sampling rates and low bitrates.
The APCodec possesses the advantages of high decoded audio quality, high compression rate, fast generation speed, low model complexity, and low latency.

\subsection{Analysis and Discussion}
\label{subsec: Analysis and Discussion}

We conducted additional analytical experiments, discussing the roles of the proposed structures and losses in APCodec through ablation studies.
Simultaneously, we explored the performance of APCodec on various other types of audio.
For simplicity, the experiments were conducted only at sampling rate of 48 kHz and bitrate of 6 kbps.

\subsubsection{Ablation Studies}
\label{subsubsec: Ablation Studies}

We conducted six ablation experiments to validate the roles of certain structures and losses in the APCodec.
The effects of other structures and losses have been confirmed in our previous publication \cite{ai2023apnet}.
For the \textbf{APCodec}, the descriptions of the ablation variants for comparison are as follows.

\begin{itemize}

\item {}{\textbf{APCodec w/o CNV}}: Ablating the modified ConvNeXt v2 network and replacing it with the residual convolutional network (RCNet) as utilized in \cite{kong2020hifi,ai2023neural,ai2023apnet}.

\item {}{\textbf{APCodec w/o MelMSE}}: Ablating the MSE loss on mel spectrograms from $\mathcal L_{M}$ in Equation \ref{equ: M Loss}.

\item {}{\textbf{APCodec w/o QLoss}}: Ablating the quantization loss $\mathcal L_{Q}$ in Equation \ref{equ: quantization Loss}.

\item {}{\textbf{APCodec w/o MRD}}: Ablating the MRD in the GAN-based loss and replacing it with the multi-scale discriminator (MSD) as utilized in \cite{kong2020hifi,ai2023apnet}.

\item {}{\textbf{APCodec w/o Hinge}}: Ablating the adversarial loss in the form of hinge GAN and adopting the one in the form of least squares GAN as utilized in \cite{kong2020hifi,ai2023apnet}.

\end{itemize}

For the \textbf{APCodec-S}, the description of the ablation variant for comparison is as follows.

\begin{itemize}

\item {}{\textbf{APCodec-S w/o KD}}: Ablating the knowledge distillation loss $\mathcal L_{KD}$, i.e., training the streamable student model directly without the guidance of the teacher model.

\end{itemize}

\begin{table}
\centering
    \caption{Objective experimental results for ablated codecs on the test sets of the VCTK dataset for sampling rate of 48 kHz and bitrate of 6 kbps. The \textbf{bold} numbers indicate the optimal results.}
    \resizebox{\linewidth}{!}{
    \begin{tabular}{c | c c c c}
        \hline
        \hline
        \multirow{2}{*}{Codec} & LSD & \multirow{2}{*}{STOI$\uparrow$} & \multirow{2}{*}{ViSQOL$\uparrow$} & $\text{AWPD}_{\text{IP}}$ \\
         & (dB)$\downarrow$ &  &  &  (rad)$\downarrow$\\
         \hline
         \hline
         \textbf{APCodec} & \textbf{0.818} & 0.875 & \textbf{4.07} & 1.68\\
         \textbf{APCodec w/o CNV} & 0.889 & 0.813 & 3.57 & 1.81\\
         \textbf{APCodec w/o MelMSE} & 0.830 & 0.830 & 3.79 & \textbf{1.65}\\
         \textbf{APCodec w/o QLoss} & 0.841 & 0.841 & 3.74 & 1.70\\
         \textbf{APCodec w/o MRD} & 0.825 & 0.874 & 3.95 & 1.70\\
         \textbf{APCodec w/o Hinge} & 0.823 & \textbf{0.879} & 3.92 & 1.67 \\
         \hline
         \hline
         \textbf{APCodec-S} & \textbf{0.835} & \textbf{0.865} & \textbf{3.93} & \textbf{1.74} \\
         \textbf{APCodec-S w/o KD} & 0.842 & 0.864 & 3.92 & 1.79 \\
        \hline
        \hline
    \end{tabular}}
\label{tab_ablation}
\end{table}

The results of the ablation experiments are shown in Table \ref{tab_ablation}.
For simplicity, only the LSD, STOI, ViSQOL and $\text{AWPD}_{\text{IP}}$ metrics were used.
By comparing the \textbf{APCodec} and \textbf{APCodec w/o CNV}, it can be observed that replacing the modified ConvNeXt v2 network with RCNet resulted in a significant decrease in all metrics.
The \textbf{APCodec w/o CNV}'s ViSQOL score decreased by 0.5 compared to the \textbf{APCodec}, indicating a significant distortion in the overall audio quality.
Specifically, according to the results of LSD and $\text{AWPD}_{\text{IP}}$, the RCNet impeded the learning of both amplitude and phase spectra, contradicting the conclusions in \cite{ai2023apnet}.
Hence, the RCNet was apt for tasks involving vocoders \cite{kong2020hifi,ai2023apnet}, leveraging its cumulative dilated convolutional layers to broaden the receptive field.
Nevertheless, in codec tasks that necessitated a more sophisticated parallel amplitude and phase modeling, the RCNet exhibited an inadequate modeling capability.
The ConvNeXt v2 network, borrowed from the field of image processing, exhibited stronger modeling capabilities, making it well-suited for the design of codec models.

Regarding the ablation studies on some training strategies, by comparing the \textbf{APCodec} and \textbf{APCodec w/o MelMSE}, it is evident that the MSE loss on the mel spectrogram had a positive impact on intelligibility and overall audio quality.
The MSE exhibited greater sensitivity to outliers when compared to MAE.
It can be viewed as a complement to MAE, collectively enhancing the overall quality of the mel spectrogram.
Derived from the outcomes of the LSD, it is reasonable that removing the amplitude-related mel-spectrogram MSE loss would lead to worse LSD.
However, metric $\text{AWPD}_{\text{IP}}$ has improved, which might be because removing the mel-spectrogram MSE loss increased the weight of the phase loss.
By comparing the \textbf{APCodec} and \textbf{APCodec w/o QLoss}, it can be observed that the quantization loss had a significant impact on the amplitude spectrum quality, intelligibility, and overall audio quality, while its influence on the phase spectrum quality was relatively minor.
The incorporation of quantization loss effectively alleviated quantization errors, thereby contributing to the enhancement of APCodec's performance.
Replacing MRD with MSD significantly impacted the amplitude spectrum quality and overall audio quality, based on the results of the \textbf{APCodec w/o MRD}.
This is in line with expectations because the MRD focused more on the quality of the amplitude spectrum, making it suitable for our spectrum-based approach.
Finally, the hinge-form adversarial loss was more effective compared to the least-squares-form adversarial loss commonly used in some vocoder tasks \cite{kong2020hifi,ai2023apnet}, according to the results of the \textbf{APCodec w/o Hinge}. 
Although replacing the adversarial loss form resulted in a 0.004 increase in STOI, this difference was not significant according to a $t$-test, while the ViSQOL value significantly decreased. 
This indicates that the hinge-form, compared to the least-squares-form form, does not improve intelligibility but significantly enhances audio quality.
In terms of auditory sensation, the \textbf{APCodec w/o Hinge} experienced very apparent harsh noise.

In terms of the role of the proposed knowledge distillation strategy for the streamable APCodec, we compared the \textbf{APCodec-S} and \textbf{APCodec-S w/o KD}.
The results are also listed in Table \ref{tab_ablation}.
It can be observed that without the guidance of the non-streamable teacher model, the streamable student model exhibited slight decreases across all metrics.
In particular, the $\text{AWPD}_{\text{IP}}$ of the \textbf{APCodec-S w/o KD} deteriorated to the level of the initial phase error, resembling the patterns seen in \textbf{Encodec}, \textbf{AudioDec} and \textbf{DAC}.
This indicates that the low-latency implementation on model structures discussed in Section \ref{subsec: Low-latency Implementation by Knowledge Distillation} hindered the learning of phase, and accurate phase prediction required a network with a broader receptive field.
The knowledge distillation strategy can effectively alleviate the challenge of phase learning difficulty ($\text{AWPD}_{\text{IP}}$ reduced by 0.05), thereby promoting overall audio quality improvement.

\begin{table}
\centering
    \caption{Objective experimental results for the comparison between \textbf{APCodec} and \textbf{DAC} and the comparison between \textbf{APCodec-S} and \textbf{AudioDec} on the test sets of the Common Voice dataset, Opencpop dataset and FSD50K dataset for sampling rate of 48 kHz and bitrate of 6 kbps, when finetuning on these three datasets individually. The \textbf{bold} numbers indicate the optimal results.}
    \begin{tabular}{c| c | c c c}
        \hline
        \hline
        \multirow{2}{*}{Dataset} & \multirow{2}{*}{Codec} & LSD & \multirow{2}{*}{ViSQOL$\uparrow$} & $\text{AWPD}_{\text{IP}}$ \\
         & & (dB)$\downarrow$ &  &  (rad)$\downarrow$\\
         \hline
         \hline
         \multirow{4}{*}{Common Voice} & \textbf{APCodec} & \textbf{0.872} & \textbf{4.25} & \textbf{1.73}\\
         & \textbf{DAC} \cite{kumar2024high} & 0.955 & 4.08 & 1.78\\
         \cline{2-5}
         & \textbf{APCodec-S} & \textbf{0.838} & 4.08 & \textbf{1.75}\\
         & \textbf{AudioDec} \cite{wu2023audiodec} & 0.929 & \textbf{4.10} & 1.80\\
         \hline
         \hline
         \multirow{4}{*}{Opencpop} & \textbf{APCodec} & \textbf{0.864} & \textbf{4.21} & \textbf{1.67}\\
         & \textbf{DAC} \cite{kumar2024high} & 0.972 & 3.94 & 1.77\\
         \cline{2-5}
         & \textbf{APCodec-S} & \textbf{0.964} & 4.06 & \textbf{1.74}\\
         & \textbf{AudioDec} \cite{wu2023audiodec} & 0.967 & \textbf{4.10} & 1.81\\
        \hline
        \hline
         \multirow{4}{*}{FSD50K} & \textbf{APCodec} & \textbf{0.853} & \textbf{3.95} & \textbf{1.71}\\
         & \textbf{DAC} \cite{kumar2024high} & 0.929 & 3.86 & 1.78\\
         \cline{2-5}
         & \textbf{APCodec-S} & \textbf{0.887} & \textbf{3.86} & \textbf{1.76}\\
         & \textbf{AudioDec} \cite{wu2023audiodec} & 1.04 & 3.77 & 1.82\\
        \hline
        \hline
    \end{tabular}
\label{tab_audio}
\end{table}


\subsubsection{Validation on Diverse Audio Datasets}
\label{subsubsec: Validation on Diverse Audio Datasets}

Since the VCTK is a small-scale speech dataset, to assess the performance of the proposed APCodec on different sizes and types of audio datasets, we incorporated three additional datasets.
These three additional datasets included: Common Voice \cite{ardila2019common}, a large-scale massively-multilingual transcribed speech corpus of approximately 919 hours; Opencpop \cite{wang2022opencpop}, a publicly available high quality Mandarin singing corpus of approximately 5.2 hours designed for singing voice synthesis; FSD50K \cite{fonseca2021fsd50k}, an open dataset of approximately 84 hours for human-labeled sound events.
For the Common Voice dataset\footnote{\url{https://commonvoice.mozilla.org/en/datasets}.}, we used the ``Common Voice Corpus 17.0" data on the download website and selected speech utterances with a sampling rate of 48 kHz.
568,822 and 6,026 utterances were respectively chosen as the training set and test set.
For the Opencpop dataset\footnote{\url{https://wenet.org.cn/opencpop/}.}, we utilized the officially pre-trimmed data, selecting 3,367 utterances as the training set and the remaining 389 utterances as the test set.
For the FSD50K dataset\footnote{\url{https://zenodo.org/records/4060432}.}, 40,966 and 4,436 utterances were respectively chosen as the training set and test set.
The sampling rates for the Opencpop and FSD50K datasets are both 44.1 kHz.
We upsampled them to 48 kHz for experiments.

For the sake of fairness and simplicity, we compared the performance between non-streamable \textbf{APCodec} and \textbf{DAC}, as well as the performance between streamable \textbf{APCodec-S} and \textbf{AudioDec}, respectively.
These models were further finetuned for 200k steps each on the Common Voice, Opencpop and FSD50K datasets, based on the well-trained models using the VCTK dataset.
We separately calculated the objective metrics for these three datasets on their respective test sets, and the results are shown in Table \ref{tab_audio}.
Due to the fact that the STOI is typically used solely for assessing speech intelligibility, we exclusively employed the LSD, ViSQOL, and $\text{AWPD}_{\text{IP}}$ metrics in this experiment.
It can be observed that, whether on the Common Voice, Opencpop or the FSD50K dataset, the performance for all metrics of the \textbf{APCodec} was significantly better than that of the \textbf{DAC}.
This confirms that our proposed APCodec still outperformed the DAC on larger speech datasets or other types of audio datasets.
When comparing the \textbf{APCodec-S} and \textbf{AudioDec} on the Common Voice and Opencpop datasets, despite the \textbf{APCodec-S} being superior in terms of amplitude and phase quality, its overall objective quality, according to ViSQOL results, was slightly inferior to that of \textbf{AudioDec}.
However, on the FSD50K dataset, all metrics of the \textbf{AudioDec} were inferior to those of the \textbf{APCodec-S}.
The above experimental results fully demonstrated that our proposed APCodec exhibited strong generalization and adaptability on other types of datasets, especially on non-human vocalization datasets, compared to other mainstream neural codecs.
Thus, the APCodec is more suitable for various audio signal processing tasks which will also be a focus of our future work.

\section{Conclusion}
\label{sec: Conclusion}

In this paper, we proposed a novel neural audio codec called APCodec.
The APCodec leveraged the advantages of parametric codecs, regarding the audio amplitude and phase spectra as parametric characteristics rather than the raw waveforms for parallel encoding and parallel decoding.
Thus, it could obtain latent codes at low frame rate using very minimal downsampling operations.
To ensure the fidelity of the decoded audio similar to waveform codecs, spectral-level loss, quantization loss, and GAN-based loss were employed to train the APCodec model.
We also constructed a low-latency streamable APCodec by combining feed-forward layers and causal deconvolutional layers with knowledge distillation training strategies.
Experimental results confirm that our proposed APCodec exhibited advantages at high waveform sampling rates and low bitrates, demonstrating high-quality decoded audio, high compression rate, fast generation speed, low model complexity, and low latency.
It surpassed the performance of the baseline Encodec, AudioDec and DAC.
Further analysis experiments also confirmed the effectiveness of the structure and loss proposed in APCodec, as well as its versatility and generalizability across diverse audio datasets.

In future work, we will 1) attempt to use features from other spectral domains, such as MDCT spectrum, as encoding and decoding objects to further enhance the training and generation efficiency of the existing framework of APCodec;
2) apply the APCodec to downstream tasks such as TTS and speech enhancement (SE), aiming to create more advanced results.

\bibliographystyle{IEEEtran}
\bibliography{mybib}
\end{document}